\documentclass[11pt]{article}
\usepackage{latexsym}
\usepackage{amsmath}
\usepackage{amssymb}
\usepackage{amsfonts}
\usepackage{epsfig}
\newcommand{\be}{\begin{equation}}
\newcommand{\ee}{\end{equation}}
\newcommand{\bea}{\begin{eqnarray}}
\newcommand{\eea}{\end{eqnarray}}

\newcommand{\la}{\leftrightarrow}

\DeclareMathSymbol{\mg}{\mathrel}{symbols}{"1D}

\newcommand{\ga}{\alpha}
\newcommand{\gb}{\beta}
\newcommand{\gd}{\delta}
\renewcommand{\ge}{\epsilon}

\newcommand{\gx}{\xi}
\newcommand{\gm}{\mu}
\newcommand{\gn}{\nu}

\newcommand{\gth}{\theta}

\newcommand{\gt}{\tau}

\newcommand{\gz}{\zeta}
\newcommand{\gp}{\pi}

\newcommand{\get}{\eta}

\newcommand{\gG}{\Gamma}
\newcommand{\gD}{\Delta}

\newcommand{\gL}{\Lambda}
\newcommand{\gS}{\Sigma}

\newcommand{\gO}{\Omega}

\newcommand{\cA}{{\cal A}}

\newcommand{\cI}{{\cal I}}
\newcommand{\cJ}{{\cal J}}
\newcommand{\cK}{{\cal K}}
\newcommand{\cL}{{\cal L}}
\newcommand{\cM}{{\cal M}}
\newcommand{\cN}{{\cal N}}
\newcommand{\cO}{{\cal O}}

\newcommand{\cR}{{\cal R}}
\newcommand{\cS}{{\cal S}}
\newcommand{\cT}{{\cal T}}

\newcommand{\ugL}{{\underline\Lambda}}

\newcommand{\bgL}{{\bar\Lambda}}

\newcommand{\ua}{{\underline a}} 
\newcommand{\ub}{{\underline b}} 
\newcommand{\uc}{{\underline c}}

\newcommand{\ug}{{\underline g}}

\newcommand{\uv}{{\underline v}}

\newcommand{\uG}{{\underline G}}

\newcommand{\ugx}{{\underline\xi}}
\newcommand{\bgx}{{\bar\xi}}

\newcommand{\bZ}{{\bf Z}}
\newcommand{\ba}{{\bar a}} 
\newcommand{\bb}{{\bar b}} 
\newcommand{\bc}{{\bar c}}

\newcommand{\bg}{{\bar g}}

\newcommand{\bv}{{\bar v}}

\newcommand{\bG}{{\bar G}}

\newcommand{\Tr}{\mbox{Tr}}
\newcommand{\tr}{\text{tr}}

\newcommand{\ra}{\rightarrow}
\renewcommand{\Re}{\text{Re}\ }

\newcommand{\inv}{^{-1}}

\newcommand{\equ}[1]{\begin{gather} #1 \end{gather}}

\newcommand{\arry}[2]{\begin{array}{#1} #2 \end{array}}

\newcommand{\pmtrx}[1]{\begin{pmatrix} #1 \end{pmatrix}}

\newcounter{oldcounter}

\if@twoside\oddsidemargin25mm
\else
\begin{document}

\begin{flushright} 
DAMTP-2002-40
\end{flushright} 
\vskip 2 cm
\begin{center}
{\Large {\bf String Threshold corrections from Field Theory}}
\\[0pt]
\bigskip
\vspace{0.7cm} 
\bigskip {\large
{\bf D.M.\ Ghilencea,$^{a,}$\footnote{
{{{\ {\ {\ E-mail: D.M.Ghilencea@damtp.cam.ac.uk}}}}}} }
{\bf S.\ Groot Nibbelink,$^{b,}$\footnote{
{{{\ {\ {\ E-mail: nibblink@th.physik.uni-bonn.de}}}}}} } 
\vspace{0.8cm}\\
{\it $^a$ DAMTP, CMS, University of Cambridge,} \\
{\it Wilberforce Road, Cambridge, CB3 0WA, United Kingdom.}\\
\vspace{0.33cm}
{\it $^b$ Physikalisches Institut der Universit\"at Bonn,} \\
{\it Nussallee 12, 53115 Bonn, Germany.}}
\end{center}
\begin{center}
\vspace{3.3cm} 
Abstract 
\end{center}
{\small{
A field theory approach to summing threshold effects 
to the gauge couplings in a two-torus compactification
is presented and the link with the (heterotic) string calculation is 
carefully investigated. We analyse whether the complete UV behaviour of 
the theory may be described on pure field theory grounds, as due 
to momentum modes only, and address the 
role of the winding modes, not included by the field theory approach. 
 ``Non-decoupling'' effects in the low energy limit, due to a 
small dimension  are discussed. The role of modular invariance 
in ensuring a finite (heterotic) string result is addressed.
}}

\newpage
\section{Introduction}
There has recently been  growing interest in the physics of 
(large) extra dimensions from a field theory perspective. 
The phenomenological implications of the associated Kaluza-Klein 
states (whether charged or not under the gauge
group) in such models have been  extensively addressed.
Many of these phenomenological studies  are performed  in an 
effective field theory approach, and the link between such Kaluza-Klein 
models and a full string theory (either heterotic or type I), where such 
effects can be consistently investigated, is not always clear.

For the purpose of the present work, we  would like to address this link 
for the particular problem  of threshold effects to the gauge couplings due to 
Kaluza-Klein (and, for closed string theory, winding modes) 
excitations associated with the
compact dimensions, in a generic orbifold compactification to 4 dimensions.
This is considered to be   a $\cN =1$ orbifold, with a (two
dimensional) $\cN =2$ sub-sector which survives compactification. The
massive momentum and winding excitations associated with the latter 
fall into $\cN =2$ multiplets and give significant corrections to the gauge 
couplings, while the massless states fall into $\cN =1$ multiplets associated
with the light spectrum and bring in a 
logarithmic contribution to the RG flow of the gauge couplings.

A field theory calculation of the threshold effects due (only) to
Kaluza-Klein states  corresponds to a zero slope limit
$\alpha'\rightarrow 0$ at string level.
In this limit string physics, such as  all ultra-heavy winding modes,  
decouple. Of course this limit is singular, and  string thresholds 
present a divergent  behaviour. Therefore it is not surprising that the
field theory result is also divergent. By comparing the two results we would like to
investigate, whether this divergence  is {\it entirely} due to Kaluza-Klein
modes, or whether winding modes can bring additional UV effects.
There are differences in the structure of the string thresholds
in the type I  case relative to the heterotic case:  in the former 
the quadratic divergence in the zero slope limit is absent, 
while the latter and field theory are sensitivity to a high cut-off scale. 
We briefly comment the origin of such behaviour.

Unlike the heterotic string case, the effective field theory approach
is a priori unable to account for  the additional effects of the winding
states which do play a role in the compactification. 
Even decoupled, winding modes may in principle still provide some additional 
(UV finite) contribution because their decoupling is, in fact, only asymptotic
\cite{Kutasov:1990sv}. Effects which may vanish 
in the limit $\alpha'\rightarrow 0$ may not be visible in a field
theory description. We identify the link between these effects  
and the mixing between momentum and winding modes.
For the type I string case  there are no winding modes
contributions to the gauge couplings, thus one may expect the finite
part be similar  to that of  regularised  effective field theory. 
Our purpose is thus to clarify for the heterotic case
which contributions of the (string) threshold correction to 
the gauge couplings cannot be computed on field theory grounds. 
We analyse the correctness of the naive expectation that their origin
be due to winding modes alone.
We also comment on the link of type I result with a field theory approach.

The plan of the work is as follows: in the next section we quote the 
heterotic and type I  string corrections to the gauge couplings,
due to momentum and winding
modes associated with the extra compact dimensions of the string. 
This will set up the notation and stage the results from the
string to which we report our findings from a  field theory approach.
In Section \ref{QFT} we use Coleman-Weinberg formula to compute the
effects of the same Kaluza-Klein spectrum to the gauge couplings. 
We take into account the effects of the shape of the manifold (two
torus), so far not considered  at field theory level.
Section \ref{QFTstring} clarifies the link field theory -
string theory and the differences in the results obtained. 
Appendix \ref{computingJ}, \ref{alternative} detail  calculations of section
\ref{QFT}, Appendix \ref{link_regularisation} - the link among the 
regularisation schemes used in the field theory  approach to computing string
thresholds, while a detailed exposition of Poisson re-summation applied to
Gaussian sums like the string partition function is given in
Appendix \ref{poissonresummation}.

\section{Thresholds results from string theory}
String threshold effects have been  computed in the heterotic 
string case in \cite{Kaplunovsky:1987rp,Dixon:1990pc,Mayr:1993mq,
Kaplunovsky:1995jw,Mayr:1995rx} 
and in type I case in \cite{Antoniadis:1999ge}. 
Before  proceeding with a field theory investigation
of these effects, we quote these well known results for later
comparison and for establishing our notations/conventions.

Kaluza-Klein states and in the case of the heterotic string also
winding states affect significantly  the value of the gauge couplings at
a scale of the order of the inverse radii of the manifold of the 
extra compact dimensions. In the case an $\cN =2$ sector ``survives'' 
compactification,  the effective gauge couplings $\alpha_i=g_i^2/(4\pi)$
receive the corrections 
\begin{equation}
\alpha _{i}^{-1}(Q) = k_{i}\alpha _{u}^{-1}+\frac{b_{i}}{2\pi } \ln
\frac{M_{s}}{Q}+\Delta _{i} +\cdots,
\label{gauge}
\end{equation}
at scale $Q$, with $k_{i}$ is the Kac-Moody level, $\Delta _{i}$ is
the (one-loop) {\it gauge group dependent} string 
threshold correction (a function of the  moduli fields) induced by $\cN=2$
multiplets. $b_i$ is the one-loop beta function of the light 
modes (${\cal N}=1$ sector). In the case of the heterotic string
a universal (gauge group independent) correction $\sigma$ exists, 
but  in (\ref{gauge}) it is absorbed  \cite{Nilles:1997vk} 
 into  the re-defined coupling $\alpha_{u}^{-1}:= \mbox{Re}\, S+\sigma$.
This definition of $\alpha_u$  is invariant under the perturbative duality
group $SL(2,Z)_T\times SL(2,Z)_U\times Z_2^{T \la U}$ of the 
heterotic string and is thus the appropriate  string expansion parameter.
It is  this coupling $\alpha_u$ which  should then be 
identified with the effective coupling at field theory level \cite{Nilles:1997vk}. 
The dots in (\ref{gauge})  stand for  higher order corrections, and 
mixing terms between (massless) twisted  states and $\cN=2$ or $\cN=4$ states.
Eq.\ (\ref{gauge}) fixes the normalisation of the thresholds $\Delta_i$
relative to the gauge couplings. 
To distinguish between the (gauge dependent part of the) 
heterotic and type I string thresholds hereafter we denote them with 
$\Delta_i^H$ and $\Delta_i^I$, respectively.

\vspace{0.5cm}
\subsection*{Heterotic string results}
\label{heterotic_results}
In this case six of the dimensions are compactified on an 
orbifold, $\cT^{6}/G$, 
and we consider the generic  case when the spectrum splits into ${\cal N}=1,$
${\cal N}=2$ and ${\cal N}=4$ sectors, with the latter two associated 
with a $\cT^{2}\times \cT^{4}$ split of the $\cT^{6}$ torus. 
Due to the supersymmetric non-renormalisation theorem, the 
${\cal N}=4$ sector does not contribute to the holomorphic
coupling, nor does it affect  the effective coupling $\alpha_i$
(\ref{gauge}) in the one-loop case. The 
${\cal N}=1$ sector brings in a  logarithmic running associated with light
states, eq.\ (\ref{gauge}).  The moduli dependence  
 brought in by  $\Delta_i$ comes  with a coefficient proportional to
the  beta function of the  ${\cal N}=2$ sector. 
All heterotic  states are closed string
states and at one loop the string world sheet has the topology of the
torus $\cT^2$. For the case of a six-dimensional supersymmetric string vacuum
compactified on a two torus $\cT^{2}$ the string correction 
\cite{Kaplunovsky:1987rp,Dixon:1990pc} (see also
\cite{Mayr:1993mq,Mayr:1995rx})  is the result 
of two double-sums over the Kaluza-Klein and winding modes, 
respectively,  associated with $\cT^{2}$,  and takes the form
\begin{equation}\label{t2form}
\Delta _{i}^{H} =-\frac{\overline{b}_{i}}{4\pi }\ln \left\{ 
\frac{8\pi e^{1-\gamma _{E}}}{3\sqrt{3}}\,U_2\,{|}\eta (U){|}^{4}\, 
T_{2}\left| \eta \left( {T}\right) \right|^{4}\right\}, 
\quad 
\eta(T)=e^{\frac {\pi}{12} iT}
\prod_{k=1}^{\infty}\left( 1-e^{2\pi ikT} \right). 
\end{equation}
Here $\overline{b}_{i}$ is the beta coefficient 
associated with the ${\cal N}=2$ sub-sector, 
and the Dedekind function $\get(T)$ is used. 
The complex moduli parameter $T = T_1 + i T_2$ and 
$U = U_1 + i U_2$ can be expressed in function of the metric $G_{ij}$
(and its determinant $G$) 
and anti-symmetric tensor $B_{ij} = B \ge_{ij}$ background as 
\begin{equation}
T = 2\Bigl[ B+ i \sqrt G\Bigr] = 
2 \Bigl[ B + i \frac {R_1 R_2 \sin \gth}{2 \ga'}\Bigr ],
\qquad 
U = \frac{1}{G_{11}}\Bigl[ G_{12}+i \sqrt G\Bigr ] 
= \frac {R_2}{R_1} e^{i \gth},
\label{moduli}
\end{equation}
with the string parameter $\alpha'$ related to the string scale $M_s$
by 
\(
M_{s}= 
{2\,e^{(1-\gamma _{E})/2}\,3^{-3/4}} / {\sqrt{2\pi \alpha ^{\prime }}}
\) 
(in ${\overline {DR}}$ scheme, see \cite{Kaplunovsky:1987rp}).  
The volume of compactification is characterized by $T_2$ ($\alpha'$
dependent), while $U$ describes the shape of the torus ($\ga'$
independent). 
In particular, for an orthogonal torus $\theta=\pi/2$ the shape parameter
$U$ is purely  imaginary. 
(Generalisations of the above result for the case of Wilson line moduli exist
\cite{Mayr:1995rx}, but for the purpose of this work we 
restrict ourselves  to the case when these are not present.)

\vspace{0.3cm}
\subsection*{Type I string results}
The threshold effects in  four dimensional $\bZ_N$ orientifold models of 
type I string with $\cN =2$ and $\cN =1$ supersymmetry have a structure
closely related to that of the heterotic string. There exists a contribution
from the twisted moduli: their VEVs remove the orbifold
singularities and allow a transition to smooth (Calabi-Yau) manifolds.
When their VEVs are set to zero by the vanishing of the D-terms of 
anomalous U(1)'s the corrections to the gauge couplings,
eq.\ (\ref{gauge}) are \cite{Antoniadis:1999ge}
\begin{equation}\label{t1form}
\Delta_i^I=-\frac{\overline{b}_{i}}{4\pi }\ln \left\{ 
{4\pi e^{-\gamma _{E}}}\,U_2\,{|}\eta (U){|}^{4}\, 
T_{2} \right\}.  
\end{equation} 
The definition of the complex structure moduli $U$ is similar to the
heterotic case, while $T$ is now expressed in type I string units
\cite{Antoniadis:1999ge}.
An additional summation (not written explicitly) is present 
if each of the three complex dimensions  comes with a $\cN =2$ sector.
The constant under the logarithm depends on the regularisation 
used at string  level\footnote{This constant is here chosen in agreement with 
the regularisation we used below for the field theory case.}.
The essential difference in (\ref{t1form}) from the heterotic
case  (\ref{t2form}) is the absence of the contribution due to
Dedekind eta function $\ln|\eta(T)|^4$, dominated by a power-like
(with the scale) term, with an additional milder (exponentially suppressed) 
contribution.

\vspace{1cm}
\section{Field theory calculation of the thresholds}
\label{QFT}
At field theory level the effects on the gauge couplings
of a finite/infinite tower of Kaluza Klein states (on a two torus) 
can be considered as well. We perform this calculation
and compare the field theory result $\Omega_i$ 
to the full string result for the gauge dependent part of the thresholds
($\Delta_i^H$ or $\Delta_i^I$). 
We will  thus identify the contribution of Kaluza Klein states only, 
and should this be possible, analyse the UV role (if any) of  
pure winding modes' contributions by comparing  to the full 
(heterotic) string result.

We start with the general Coleman Weinberg  formula for the threshold
effects \cite{Kaplunovsky:1987rp} to the gauge couplings. We have
\begin{equation}\label{QFTthresholds1}
\Omega_{i} = \frac{1}{4\pi}\sum_{i} T(R_i) 
\sum_{m_{1,2} \in \bZ}' \int_{\xi}^{\infty}\frac{dt}{t} 
\, e^{-\pi\,t M_{m_1,m_2}^2/\mu^2 }.  
\end{equation}
The result of summing a finite or infinite tower of Kaluza-Klein 
states to the gauge couplings is divergent, thus the need for a 
regularisation.  We introduced a UV dimensionless (proper time) regulator 
$\xi\ra 0$ as the lower limit of the integral above. This scheme will
be used in the following.  Other regularisations can be used, with 
{\it formal} equations relating them
presented in Appendix \ref{link_regularisation}. 
An alternative method to computing (\ref{QFTthresholds1}) is given 
in Appendix \ref{alternative},  using dimensional reduction 
or zeta function regularisation 
scheme.
This scheme is {\it explicitly} 
related to the cut-off regularisation that we use in this section.
The method used in Appendix \ref{alternative} clarifies the link
between these regularisations, to relate results obtained
in such different schemes. This method is 
general and may be applied to other calculations as well (e.g. 
the scalar potential in Kaluza Klein models).
Finally, we mention that in all regularisations employed in this work,
the Clifford algebra is worked out first,  and only after is the
regularisation of the scalar sum-integrals of type
(\ref{QFTthresholds1}) performed.  Therefore, all supersymmetric 
cancellations that rely on spinor properties are taken into account,
before the regularisation of $\Omega_i$. 

In (\ref{QFTthresholds1}) $\mu$ is an arbitrary (finite) mass scale, 
which we introduced to render the above equation dimensionless.
A ``prime'' on the sum  in eq.\ (\ref{QFTthresholds1}) stands for the 
absence of  the  $(m_1,m_2)=(0,0)$ Kaluza Klein massless mode 
which at this stage is subtracted out. 
This is because this ($\cN =1$)  massless mode is accounted for by the
logarithm of the cut-off already present in eq.\ (\ref{gauge}).
In (\ref{QFTthresholds1}) $M^2_{m_1,m_2}$ is  the mass expression for 
Kaluza-Klein states for a two dimensional case (torus), with ``shape
effects'' of the torus characterised by the ratio of radii {\it and} 
angle $\theta$.  These  together define the quantity $U$ 
(with definition as in string case eq.\ (\ref{moduli}) since it is scale
independent) and the volume $T$ 
\begin{equation}\label{fieldcase}
{T} (\mu) \equiv i T_2 (\mu)= i \mu^2 R_1 R_2 \sin\theta\, .
\end{equation}
Definition (\ref{fieldcase}) for $T_2$ will only 
be used in the field theory case, instead of its string version (\ref{moduli}).
Thus the mass of Kaluza-Klein states on a two-torus can be written in terms of
$U$ and $T$ as (for details  see  e.g. \cite{Dienes:2001wu}) 
\begin{equation}\label{kkmass}
M^2_{m_1,m_2}=
\frac{1}{\sin^2\theta}\left[\,\frac{m_1^2}{R_1^2}+\frac{m_2^2}{R_2^2}-
\frac{2 m_1 m_2 \cos\theta}{R_1 R_2}\,\right]
\,\, =\,\, \frac{\vert m_2-U m_1\vert^2 }{ (\mu^{-2} T_2) U_2} .
\end{equation}
In the limiting case  $\theta=\pi/2$,  
$M^2_{m_1,m_2} \rightarrow {m_1^2}/{R_1^2}+{m_2^2}/{R_2^2}$
which is valid for an orthogonal torus, more commonly encountered in
the literature than the general case of arbitrary $\theta$ that we consider here. 
Eq.\ (\ref{kkmass}) shows that even for large compactification
radii $R_{1,2}$, the Kaluza-Klein states may have very large mass and thus 
be undetectable if $\theta\ll 1 $  as may be seen \cite{Dienes:2001wu} by
expanding (\ref{kkmass}) for small $\theta$.

The integrand  of (\ref{QFTthresholds1})  may be written as 
\begin{eqnarray}\label{cI1}
\cI&\equiv&\!\!\sum_{m_{1,2} \in \bZ}' \! 
e^{-\frac{\pi\,t}{T_2 U_2} |U m_1-m_2|^2} 
=\sum_{m_{2} \in \bZ}' \, e^{-\frac{\pi\,t}{T_2 U_2} m_2^2} 
+\sum_{m_{1} \in \bZ}' \sum_{m_2 \in \bZ}\, 
e^{-\frac{\pi\,t}{T_2 U_2}  |U m_1-m_2|^2}\\
\nonumber\\
&=&\!\!\!
\sum_{m_{2} \in \bZ}' \! e^{-\frac{\pi\,t}{T_2 U_2} m_2^2} 
+\bigg[\frac{T_2 U_2}{t}\bigg]^{\frac{1}{2}}\!\!\!
\sum_{m_{1} \in \bZ}' \! e^{-{\pi\,t}\frac{U_2}{T_2} m_1^2}
+\bigg[\frac{T_2 U_2}{t}\bigg]^{\frac{1}{2}}\!\!\!
\sum_{m_{1} \in \bZ}'\sum_{p \in \bZ}' \!
e^{-{\pi\,t  \frac{U_2}{T_2} m_1^2 -\frac{\pi}{t} T_2 U_2\, p^2
-2 i \pi m_1 p\, U_2}}, 
\label{cI2}
\end{eqnarray}
with $m_2$ replaced in (\ref{cI2}) by the Poisson re-summed index $p$, 
using eq.\ (\ref{poisson1}).
Each of the sums above  can be integrated over  $t\in [\xi,\infty)$ separately.
From (\ref{QFTthresholds1}), (\ref{cI2}) we then have  
\begin{equation}\label{omegai}
\Omega_i=\frac{{\overline {b}}_i}{4\pi}\left(\cJ_1+\cJ_2+\cJ_3\right),
\end{equation}
with ${\overline b_i}= - 2 T(G)+2 \sum_{r} T_i (r)$ the $\cN =2$ beta
function  and  with (see  Appendix (\ref{s1}) and (\ref{r1})) 
\begin{eqnarray}
\cJ_1&\equiv&\int_{\xi}^{\infty} \frac{dt}{t}
\sum_{m_{2} \in \bZ}' \, e^{-\frac{\pi\,t}{T_2 U_2} m_2^2} 
=-\ln\bigg[4 \pi e^{-\gamma_E} \, U_2
\frac{T_2}{\xi}\bigg]+2\bigg[\frac{T_2 U_2}{\xi}\bigg]^\frac{1}{2},
\label{dg1}\\
\nonumber\\
\cJ_2&\equiv&\int_{\xi}^{\infty} \frac{dt}{t}
\bigg[\frac{T_2 U_2}{t}\bigg]^{\frac{1}{2}}
\sum_{m_{1} \in \bZ}' \, e^{-{\pi\,t}\frac{U_2}{T_2} m_1^2}
=\frac{\pi}{3} U_2 +\frac{T_2}{\xi}-2 \bigg[\frac{T_2
U_2}{\xi}\bigg]^\frac{1}{2},
\label{dg2}\\
\nonumber\\
\cJ_3&\equiv&\int_{\xi}^{\infty} \frac{dt}{t}
\bigg[\frac{T_2 U_2}{t}\bigg]^\frac{1}{2}
\sum_{m_{1} \in \bZ}'\sum_{p \in \bZ}'
e^{-{\pi\,t}\frac{U_2}{T_2} m_1^2-\frac{\pi}{t} T_2 U_2 p^2}
e^{-2 i \pi m_1 p U_2}\label{dg3}\\
\nonumber\\
&=&\!\!\!  4\, U_2^{\frac{1}{2}}\!
\sum_{m_{1}\geq 1} \sum_{p\geq 1}
\bigg[\frac{m_1}{p}\bigg]^{\frac{1}{2}}\!
\Big(e^{-2 i \pi m_1 p U_1}+c.c.\Big)
K_{-\frac{1}{2}}\! \left(2 \pi U_2 m_1 p\right)\!
=-\ln \! \! \prod_{m_1\geq 1}\left\vert 1-e^{2 i \pi m_1 U}
\right\vert^4.
\label{dg4}
\end{eqnarray}
In (\ref{dg3}) the regulator $\xi$ was removed since the
integral was well defined in both UV and IR. In fact, all expressions
of $\cJ_i$ (i=1,2,3) after the (second) equal sign 
are only true in the limit 
$\xi \rightarrow 0$, and  we will  comment on the size of the 
errors induced for $\xi \neq 0$ by using the results of Appendix 
\ref{T2U2conditions}. In (\ref{dg4}) we used
the integral representation of  the modified Bessel function 
$K_\nu$ \cite{Gradshteyn} of index $\nu=1/2$.

We  identified the sources of all (UV) divergences in $\Omega_i$ 
as  coming from  $\cJ_{1}$ and $\cJ_2$. These are
logarithmic in scale, $\ln(T_2/\xi)$,  
due to ``original''  momentum ``zero modes'' 
(see eqs.\ (\ref{qq1}) to (\ref{lastline}))  and 
quadratic in scale ($T_2/\xi$) which are due to Poisson re-summed 
momentum ``zero  modes'' (see eqs.\ (\ref{sr1}) to (\ref{r3})). The linear
divergence $(T_2/\xi)^{1/2}$ in (\ref{dg1}) and (\ref{dg2}) cancels out.
The above identification will be  useful for  comparing to  the heterotic
case, where the same states will be integrated over a different 
region (the fundamental domain) and this
will avoid the divergent   behaviour of field theory.

From (\ref{QFTthresholds1}), (\ref{omegai}), 
(\ref{dg1}), (\ref{dg2}), (\ref{dg4}) we find\footnote{
Note that the non-holomorphic correction $\ln(T_2 U_2)$ 
in eq.\ (\ref{dg1}), which propagates  into the final result for $\Omega_i$,
is essentially due to a massless state  $(m_1,m_2)=(0,0)$.
This may be observed by a careful examination of 
eqs.\ (\ref{cI1}), (\ref{cI2}), (\ref{dg1}), 
(\ref{JJ1}) and also (\ref{qq1}) to (\ref{lastline}) showing the
origin of the emergent $\ln(T_2 U_2)$. 
For additional details on the non-holomorphic contribution
at string level see \cite{Kaplunovsky:1995jw}.}
\begin{equation}\label{result}
\Omega_i= - \frac{{\overline {b}}_i}{4 \pi} 
\ln\Big[4 \pi e^{-\gamma_E} \, e^{-{T_2^*}} \, {T_2^*} \, U_2
\, \vert \eta(U)\vert^4\Big], 
\end{equation}
where we introduced the notation
\begin{equation}\label{cutoff}
T_2^*\equiv\left.\, \frac{T_2}{\xi}\, \right\vert_{\xi\ra 0}
= \Lambda^2  R_1 R_2 \sin\theta, 
\qquad \Lambda^2\equiv\frac{\mu^2}{\xi}. 
\end{equation}
Eq.\ (\ref{result}) is the main result of this work. It describes the
effects  of the tower of Kaluza Klein states alone, and it includes shape
moduli effects (U  dependence) not considered by previous field 
theory approaches \cite{Dienes:1998vg}. They have  important effects
as we discuss in the next section.

The above field theory result holds true if the (dimensionless) cut-off
$\xi\ra 0$  or equivalently if the (dimensionful) cut-off scale 
$\Lambda$ introduced in (\ref{cutoff})  satisfies $\Lambda\ra \infty$.
Indeed, $\cJ_i$ (i=1,2,3) of (\ref{dg1}), (\ref{dg2}) and (\ref{dg3}) have
corrections $\delta_i$, evaluated in Appendix \ref{T2U2conditions}, eqs.\ (\ref{dj1}),  
(\ref{dj2}), (\ref{dj3}). The sum of these corrections vanishes only if $\xi\ra 0$. 
This requirement may actually be ``relaxed'' and  the following sufficient
condition for the field theory  cut-off (or radii) exists, derived in
eqs.\ (\ref{d1}), (\ref{d2}), (\ref{d3}) 
\begin{equation}\label{cond1}
T_2^* \, U_2 \gg \max\bigg\{ U_2^2, \frac{1}{U_2^2}\bigg\}. 
\end{equation}
If this condition is not satisfied, corrections $\delta_i$ to the
field theory result (\ref{result}) may
become  significant. Using  $T_2^*$ and $U_2$ definitions,  this
translates into
\begin{eqnarray}
&\Lambda \,R_2 \sin\theta > \Lambda \, R_1  \gg 1,& \qquad (U_2>1), 
\label{co1}\\
\nonumber\\
&\Lambda\, R_1 > \Lambda \, R_2 \sin \theta \gg (\Lambda
R_1)^{\frac{1}{2}},&  \qquad (U_2<1),
 \label{co2}
\end{eqnarray}
which lead to a large compactification volume  $T_2^*\equiv \gL^2 R_1 R_2
\sin \gth \gg 1$. 
While the constraint for the radius be larger than $1/\Lambda$ 
 is not surprising, one notices the role that the
angle $\theta$ plays in the validity of the field theory result, 
eq.\ (\ref{result}). The constraints are important for 
very small angle $\theta$, and have implications for the validity of
phenomenological studies with more than one dimension.

The field theory result  (\ref{result}) is very close to the heterotic 
string result of eq.\ (\ref{t2form}) and it shows quadratic and logarithmic
divergences. A comparison with the (heterotic) string result 
(\ref{t2form})   shows that the field 
theory cut-off is replaced at string level by  
$\Lambda^2 \equiv \mu^2/\xi\ra 1/\alpha'$. The condition of removing the
regulator, $\xi\ra 0$  in (\ref{QFTthresholds1}) 
corresponds to an infinite string scale limit $\alpha'\ra 0$. 
String  effects which vanish in this limit may not
necessarily be recovered by a field theory approach and these will be
 identified later on. When  $\alpha'\ra 0$  the string result
(\ref{t2form}) also shows
quadratic and logarithmic divergences,  similar to (\ref{result}). This
is a good  check of our calculation and shows the two results
are indeed very similar in this limit. One notices however
different coefficients  for the quadratic divergence 
terms,   $T_2^*$ in $\Omega_i$ and $(\pi/3)\, T_2$ in $\Delta^H$.
A suitable re-definition of the regulator $\xi$, $\xi\ra (3/\pi)\, \xi$
may avoid this difference. This may be
regarded as the correct (re)definition for a cut-off of the field theory,
such that the  UV behaviour is identical to that from string theory.
The conclusion is that in the absence of a full string calculation to
compare with, one  cannot claim that the {\it whole UV behaviour} of 
a Kaluza Klein  model is that found on pure field theory
grounds. While the nature of the divergence (quadratic) 
is correctly reproduced, 
its coefficient is not that of string theory. For this 
a full string calculation is necessary. 
We comment on this  issue in  Section \ref{analysis}.

\subsection{``Non-decoupling''  effects of a small  dimension.}\label{limits}
In this section we discuss the contribution  of ``shape effects'' ($U$
dependence) in (\ref{result}) versus ``volume'' effects 
($T$ dependence).

Consider first the case\footnote{We do not address here the
implications of large moduli from  string theory point of view. In
particular large $T_2$ may imply (for the heterotic string)
a large 10D string coupling and the reaching of its non-perturbative limit.}
 $U_2\gg 1$. Since $U_2 \equiv R_2/R_1 \sin\theta$,
this condition may be respected if for example  $\theta=\pi/2$ and
$R_2\gg R_1$. In such case, using the
definition     of the Dedekind function (\ref{t2form}) and its
asymptotic behaviour,   we find that the
leading  contribution to  $\Omega_i$ due to shape effects  
is  significant, $\Omega_i\sim (\pi/3) \, U_2\gg 1$.  
We compare this contribution to ``volume'' effects. The 
leading contribution in $T$ to 
$\Omega_i$ is $\Omega_i\sim T_2^*$ which is  larger
 than $(\pi/3) \, U_2$ because
$T_2^* \gg U_2$ for $U_2>1$, if eq.\ (\ref{cond1}) is respected. 
Therefore  ``volume'' effects are more important than ``shape'' effects. 
However, the latter are still
significant and do not necessarily decouple, although the manifold is 
effectively one-dimensional, $R_2\gg R_1$. This can have important 
phenomenological implications.

Consider now the case $U_2\ll 1$. This is possible if for example 
$R_1=R_2$ and $\theta\ll 1$. In this case one finds that the $U$
dependence of the thresholds is $\Omega_i\sim  \pi/ (3\theta)+ \ln
\theta$ which is again large (but still smaller than
volume effects. This is because the leading ``volume'' part is again
$(\pi/3) T_2^*>\pi/ (3\theta)$ because  $T_2^*\gg 1/U_2\approx 1/\theta$
for small $\theta$, eq.\ (\ref{cond1})).

One reason why shape effects do not ``decouple'' when   $R_2\gg R_1$  is that 
$U$ is scale independent. A second reason is that leading $U_2$
dependence\footnote{This argument applies for $U_2\gg 1$. For $U_2\ll
1$ see $\cJ_3$ of eq.\ (\ref{dg4}).} of
$\Omega_i$ -  similarly to its $T_2$ dependence - arises from (\ref{dg2}) as 
an effect of  ``mixing''  of states associated each with one of the 
two dimensions. Indeed $\cJ_2$ is the
result of one tower of momentum states (sum over $m_1$) due to one
dimension, combined with that of the second dimension, the
Poisson re-summed ``zero'' state ($p=0$) in $\cJ_2$ (see (\ref{cI2}) from
which  $\cJ_2$  arises). 
A final reason for the non-decoupling of $U$ dependence is that
it is due to  contributions of   {\it infinitely} many
states associated with the small(er) dimension $R_1$,  therefore their
overall effect does not necessarily  decouple.
As a further illustration of this point consider the following case with
$U_2>1$
\begin{equation}
R_2 \gg R_1\equiv   \frac{k}{\Lambda \sin\theta}\gg
\frac{1}{\Lambda},\qquad k=\cO(1). 
\label{radii}
\end{equation}
The last inequality is due to (\ref{co1}) and requires
small $\theta$, $\theta\ll 1$  which together with the assumption
 $U_2>1$ gives $R_2\gg R_1$ of eq.\ (\ref{radii}). Given this relation
between the radii, the case is ``effectively'' one dimensional.
The leading behaviour of $\Omega_i$ of (\ref{result}) is then
\begin{equation}\label{onecodim}
\Omega_i =-\ln \frac{\overline b_i}{4\pi}\bigg[4 \pi e^{-\gamma_E} \,
e^{-k \Lambda R_2}  (\Lambda R_2 \sin \theta)^2\, \left\vert \eta(R_2/R_1
e^{i \theta})\right\vert^4\bigg]
\sim \frac{\overline b_i}{4\pi} \, (k \Lambda R_2)
\end{equation}
and a linear divergence emerges, as expected for an effectively
one-dimensional case, $R_2\gg R_1$. This result should be compared to 
that for  one extra dimension ($R_2$) only. 
The contribution to the gauge thresholds in
the  one dimensional case (\ref{onedim}) is
\begin{equation}\label{onedimension}
\Omega_i^{(1)} =- \frac{\overline b_i}{4\pi}
\ln\Big[ 4\pi e^{-\gamma_E}   \, e^{- 2 \Lambda R}\, (\Lambda R)^2  \Big]
\sim \frac{\overline b_i}{4\pi} \, (2\Lambda R_2). 
\end{equation}
Comparing (\ref{onedimension}) and (\ref{onecodim})
we conclude that the coefficient of the leading (linear) 
ultraviolet divergence is in general  different in the two cases, 
unless a very specific choice is made, ($k=2$).
Therefore, in addition to the (sub-dominant) shape effects which do
not  decouple in the limit $R_2\gg R_1$, we see the same is true for
the more important UV contribution $\Lambda R_2$, 
which has a different coefficient. The
difference is essentially  due to additional (infinitely many)
states of large  mass $\sim 1/R_1 \gg 1/R_2$, c.f. (\ref{radii}),
included in (\ref{onecodim}) but not present in (\ref{onedimension}),
and  which control the value of this
UV (linear) divergence coefficient.

In the light of this discussion, applied to the  two-torus compactification
of the previous section, we can say (from a field theory
point of view) that the  coefficient (equal to 1)
of  the quadratic divergence in 
(\ref{result}) is different from that equal to $\pi/3$ of the
(heterotic)  string case (\ref{t2form}) due to additional 
states of large mass. Such states exist in the string case
and  may contribute, and were  not considered in 
eq.\ (\ref{QFTthresholds1})  (winding modes).

\section{The link with string theory}\label{QFTstring}

Since  the field theory result (\ref{result})  is close to both the
heterotic and type I string results, eqs.\ (\ref{t2form}),
(\ref{t1form}), a comparison of these results is performed
in the following. To do this we first review the (heterotic) string calculation. 
We then point out the similarities and differences with our field 
theory calculation.

\subsection{Heterotic String Case}
In this subsection  we review the  standard calculation of the heterotic 
string \cite{Dixon:1990pc} leading to (\ref{t2form}).
We consider a six-dimensional supersymmetric string vacuum
compactified on a two torus $\cT^{2}$; the remaining dimensions are 
associated with higher amount of supersymmetry and do not contribute
to the holomorphic coupling.  The string correction is 
\cite{Kaplunovsky:1987rp,Dixon:1990pc}
\begin{equation}
\Delta _{i}=\frac{\overline{b}_{i}}{4\pi }\int\limits_{\Gamma }
\frac{d\tau_{1}d\tau _{2}}{\tau _{2}}\left( Z_{torus}-1\right),  
\label{gag}
\end{equation}
where $-1$ subtracts out the contribution of massless modes 
included separately in (\ref{gauge}). The fundamental domain
is defined by $\Gamma =\{\tau _{2}>0,|\tau _{1}|<1/2,|\tau |>1\}$ 
and $\tau =\tau _{1}+i\tau_{2}$ is the modulus of the world 
sheet torus. The usual partition function in terms of 
winding modes ($n_{1,2}$) and Kaluza-Klein modes ($m_{1,2}$)
is (for an additional  discussion see  Appendix \ref{poissonresummation})
\bea\label{stringthreshold}
Z_{torus}&=&\sum_{n_{1,2},m_{1,2}\in \bZ}\exp\left[ 2\pi i\tau
(m_{1}n_{1}+m_{2}n_{2})\right] \exp\left[ -\frac{\pi \tau _{2}}{T_{2}U_{2}}
|TUn_{2}+Tn_{1}-Um_{1}+m_{2}|^{2}\right]  \\
&=&\frac{T_2}{\tau_2} \sum_{A} e^{-2\pi i T \det A}\,
\exp{\left[-\frac{\pi T_2}{\tau_2 U_2} 
\left\vert
\pmtrx{1 & U} A \pmtrx{ \tau \\  1} 
\right\vert^2\right]},
\qquad A =\pmtrx{ n_1 & p_1\\ n_2 & p_2 }, 
\label{second}
\eea
with the sum over $2 \times 2$ matrices $A$ with integer elements. 
In the last step a Poisson re-summation over the original 
Kaluza-Klein modes $m_{1,2}$ was 
performed, $m_{1,2}\rightarrow p_{1,2}$,  
to give  two double sums\footnote{Throughout this
work we make a distinction between original momentum modes $m_{1,2}$ and 
the Poisson re-summed ``modes'', $p_{1,2}$. In particular $p_{1,2}=0$ may
correspond to a large number of original Kaluza-Klein states.} over  $p_{1,2}$
and original winding modes ($n_{1,2}$).

Following eqs.\ (\ref{gag}) and (\ref{second}), 
the heterotic string  result can be written
as a sum over the orbits of the modular group $SL(2,\bZ)$ in the space
of $GL(2,\bZ)$ \cite{Dixon:1990pc} to give
\begin{equation}\label{sum}
\Delta_i=\frac{{\overline {b}}_i}{4 \pi}\left[
\cJ ^{(A=0)}+\cJ ^{(\det A=0)}+\cJ ^{(\det A\not=0)} +\int_\Gamma
\frac{d\tau_{1}d\tau _{2}}{\tau _{2}} (-1) \right].
\end{equation}
The first contribution (zero orbit) is due to (Poisson re-summed)
Kaluza-Klein modes only with  $p_{1,2}=0$ (no winding modes,
$n_{1,2}=0$),  which may be due to infinitely many original
Kaluza-Klein modes $m_{1,2}$.  The second contribution 
$\cJ^{(\det A=0)}$ (degenerate orbit) and the third $\cJ ^{(\det
A\not=0)}$ (non-degenerate orbit)
mix both windings ($n_{1,2}$) and original momentum $m_{1,2}$ (or 
equivalently $p_{1,2}$) modes. 
Finally the integral of the  last term accounts for the (field theory)
massless states ($m_{1,2}=n_{1,2}=0$) contribution.
The first contribution obtained after integrating over the fundamental
domain $\Gamma$ is finite
\begin{equation}\label{piover3}
\cJ ^{(A=0)}=
\int\limits_{\Gamma }
\frac{d\tau_{1}d\tau _{2}}{\tau _{2}^2}\, T_2=
\frac{\pi}{3} T_2.  
\end{equation}
The domain of integration $\Gamma$ ensures 
the heterotic string gives a finite result. 

The degenerate orbit contribution mixes both winding and Kaluza-Klein
modes. An ``unfolding'' procedure of the fundamental domain \cite{O'Brien:pn}
is then used to re-write the four sums over (Poisson re-summed, $p_{1,2}$) 
Kaluza-Klein and winding ($n_{1,2}$)  modes as
two sums over a mixture of these modes, while the remaining two sums 
``conspire'' with the fundamental domain of integration to give instead an
integral over the half strip  only, $(\tau_2>0, -1/2<\tau_1<1/2)$. 
The integral over $\tau_1$ is then trivial to give \cite{Dixon:1990pc}
\begin{equation}\label{detA=0}
\cJ ^{(\det A=0)}=T_2 \int_{0}^{\infty}\frac{d\tau_2}{\tau_2^2}
\sum_{j,p\in \bZ}' e^{-\frac{\pi}{\tau_2} \frac{T_2}{U_2} |j+U p|^2}
\left[ 1-e^{-\frac{N}{\tau_2}}\right] = \ln N -  \ln \Bigl[
  4 \pi  e^{-2\gamma_E} T_2 \, U_2 \,\, | \get(U) |^4 \,\,
\Bigr],
\end{equation}
Note that $j$, $p$ above are not Kaluza-Klein levels, but a mixture of both 
original winding and (Poisson re-summed, $p_{1,2}$)  momentum modes.  
In the decoupling limit of winding modes $\alpha'\rightarrow 0$, $j,p$
are mainly (Poisson re-summed) Kaluza-Klein  ``levels''. To evaluate 
(\ref{detA=0}) an infrared (IR) regulator $(1- e^{(-N/\tau_2)})$ is 
introduced ($N\ra \infty$) since  $\cJ^{(\det A=0)}$ has a divergence
coming from the IR integration region $(\tau_2\ra \infty$).
This IR divergence  cancels out upon subtracting out  the (regularised) massless
($m_{1,2}=n_{1,2}=0$) contribution, see (\ref{gag})
\begin{equation}\label{massless}
\int\limits_{\Gamma }
\frac{d\tau_{1}d\tau _{2}}{\tau _{2}} (-1)\left[1-e^{-\frac{N}{\tau_2}}\right]
=-\ln N -\left[\gamma_E+1-\ln{3\sqrt 3}/{2}\right]
\end{equation}
There remains  the contribution $\cJ^{(\det A\not=0)}$ whose
calculation leads to \cite{Dixon:1990pc}
\begin{equation}\label{windingmode}
\cJ ^{(\det A\not=0)}=
-\ln \prod_{n_1>0} \left| 1-e^{2 \pi i \,n_1\, T }\right|^4, 
\end{equation}
with $T$ expressed as usual in $\alpha'$ units.
$\cJ^{(\det A\not=0)}$  vanishes in the regime  of Kaluza-Klein modes  only,  
of $\alpha'\rightarrow 0$, see eqs.\ (\ref{second}) with $T_2
\propto 1/\alpha'$. This contribution is in essence due to the phase factor in 
(\ref{stringthreshold}) and is  zero  if $n_{1,2}=0$ (no winding modes). 
More exactly, (\ref{windingmode}) is the result of a
mixed contribution (re-summed) momentum modes - winding modes, see
definition of $A$, eq.\ (\ref{second}).
Adding together (\ref{piover3}), (\ref{detA=0}), (\ref{massless}) and 
(\ref{windingmode})  gives the string result, eq.\ (\ref{t2form}).
This ends our review of the heterotic string calculation.

\subsubsection{Analysis - Heterotic string versus field theory calculation}
\label{analysis}

We can now address  the similarities and differences
between the (heterotic) string and field theory results.

The term $(\pi/3)\, T_2\,$ of eq.\ \eqref{piover3} (zero orbit)   
gives in the regime of Kaluza Klein modes $\alpha'\ra 0$ ($T_2$
large) when  winding modes are asymptotically decoupled, the leading 
contribution to the string thresholds. 
This  term corresponds at field theory level to the quadratic  
divergence of  eq.\ (\ref{dg2}), (\ref{result})
as a result of integrating  there from $\xi\ra 0$.
Thus the domain of integration $\Gamma$ leading to  \eqref{piover3} ensures 
a finite (heterotic) string result,  while at field theory level a 
regularisation is required. 
Further, the coefficient $\pi/3$ in (\ref{piover3})  is itself
a consequence of integrating over $\Gamma$. 
Since $\Gamma$ has no field theory correspondent, a $\pi/3$ coefficient
for its quadratic divergence
is not recoverable on pure field theory grounds. 
In field theory  this coefficient is regularisation scheme dependent, 
but may be {\it chosen} to be equal to that in heterotic string, by 
a suitable regulator re-definition in eq.\ (\ref{result}),
 $\xi\ra (3/\pi) \, \xi$.  
Note that the leading (quadratic) UV behaviour
is    both in string and field theory due to Poisson re-summed  
Kaluza-Klein zero modes, $p_{1,2}=0$, eqs.\ (\ref{piover3}) and (\ref{dg2}).

The above arguments  do not imply that winding modes 
do not have any role in the UV. They only mean that 
in the regime of $\alpha'  \ra 0$ of field theory 
with a suitable choice or redefinition of the regulator, 
the UV behaviour  of the couplings can be  
described (in the heterotic case) 
on pure field theory grounds only, eq.\ (\ref{result}). 
Winding modes enable the symmetries of the
string (modular invariance) and require that the $\tau$
integration in (\ref{piover3}) take place over the fundamental domain 
$\Gamma$. 
They thus  (indirectly) ``control'' the value  of the 
resulting coefficient $\pi/3$ of (\ref{piover3}), 
not recovered on pure field theory grounds.

In the (field theory) limit of vanishing  $\alpha'$ 
the result (\ref{detA=0}) (degenerate orbits) is  due to Kaluza-Klein
modes only, having in this limit an UV logarithmic divergence, 
$\ln T_2$ with field theory correspondent 
in the UV logarithmic divergence $\ln\xi$ of (\ref{dg1}), (\ref{result}).
The remaining finite part is similar in both cases. For example 
the whole dependence on $U$ moduli
($\ln U_2 |\eta(U)|^4$)  is re-obtained by the field theory calculation.
This may be explained by the fact that $U$  is (scale)
$\alpha'$  independent. This enables  one to take account at field
theory level of the $U$ dependence, as we did in 
eq.\ (\ref{kkmass}) for the  mass spectrum of Kaluza Klein states.
This spectrum displays the symmetry  $U\ra 1/U$ (or 
equivalently $R_{1,2}\ra R_{2,1},\,  \, \theta\ra - \theta$),
and is not surprising that the  $U$ dependent part of the 
final result (\ref{result})
has (unlike the $T$ dependent part) this symmetry as well,
just like the string.
Also note that ignoring  ``zero modes'' contributions,
eq.\ (\ref{detA=0}) is equal after double Poisson re-summation 
(\ref{poisson2}) to the field theory
integrand, see eqs.\ (\ref{QFTthresholds1}), (\ref{kkmass}). Finally, the
result (\ref{detA=0}) has some winding modes' effects included, recovered on 
field theory grounds simply because the integration limit
in (\ref{QFTthresholds1}), $(\xi,\infty)$ recovers for $\xi\ra 0$ the
integration range $(0,\infty)$ of (\ref{detA=0}). In (\ref{detA=0})
this is the result of an ``unfolding'' effect of 
(additional) summing over winding modes. Taking $\xi\ra 0$
thus ``recovers'' some winding modes effects on field theory grounds,
but then the field theory calculation requires an UV regularisation.

In string theory  winding modes are needed
in deriving (\ref{detA=0}) to unfold the fundamental domain to 
the integration range $\tau_2 >0 $ and they ensure a finite 
result in the UV ($\tau_2\ra 0$). The result (\ref{detA=0}) has
an infrared  IR divergence, $\ln N$, well known 
in string theory, coming from 
integration in the region $\tau_2\ra \infty$ in eq.\ (\ref{detA=0}).
Thus the sum over all modular orbits, given by the first three terms
in (\ref{sum})  is IR divergent as well. It is also  invariant under 
modular transformations for $\tau$. Therefore the 
string corrections to the gauge coupling, although respect this
symmetry, have a divergence essentially due to  massless (field
theory) modes contribution.  Further, this  divergence ($\ln N$) is 
subtracted out  (see eq.\ (\ref{massless}))  
so that  $\Delta^H$ computes only (massive modes') corrections to
the field theory contribution, with the latter introduced 
separately  in (\ref{gauge}), the logarithmic term. (This last 
contribution is not modular invariant).
Modular invariance  does not ``forbid''  an IR divergence at string 
level  which needs to be regularised. It is not clear to us to
what  extent the fact that the regulators used  (for 
degenerate orbits integral) 
in string theory, are not invariant  under this symmetry, should be a matter
for concern, given that this symmetry is used at an earlier stage 
in ``unfolding'' and computing  the same integral(s).

Finally we observe that  at field theory level one only required
an UV regularisation for the contribution of the massive KK modes, 
while at string level an IR regularisation was instead needed.
The regularisation constant in field theory case $(-\ln(4 \pi
e^{-\gamma}))$  of (\ref{result}) emerged from (\ref{dg1}),
(\ref{ll2}), (\ref{lll2}) where zeta function/DR regularisation was 
used. This constant  is equal (see Appendix A in \cite{Foerger:1998kw}) to 
that emerging at heterotic string from eq.(\ref{detA=0}),
provided the same regularisation (DR) 
is performed in (\ref{detA=0}) in the IR regime. 
This implies a possible relationship between
the UV and IR regimes of the field theory  (which at string level is 
enabled by winding modes).

The contribution from the non-degenerate orbits (\ref{windingmode}) 
is the  only part  missed by the field theory result, eq.\
(\ref{result}) and is  related to the presence in string case
of the phase factors in (\ref{stringthreshold}), (\ref{second})
not present at field theory level. Indeed  the  Coleman Weinberg formula
eq.\ (\ref{QFTthresholds1}) is similar to the string partition function
only if $n_{1,2}=0$. 
The dependence on $T_1= \Re T= 2B$ in (\ref{windingmode})  
where $B$ is related  to the (world sheet) antisymmetric tensor 
field does  not have a (direct) field theory explanation either.
In fact (\ref{windingmode})  is a  non-perturbative correction, 
of topological nature (world sheet instantons), as it corresponds to
a mixed contribution (re-summed) Kaluza-Klein - winding modes, thus 
unrecoverable on field theory grounds. 
It is interesting to note that a term of structure similar  to
(\ref{windingmode}) but with $T\rightarrow U$ could however be computed 
by field theory, see eq.\ (\ref{result}). 
Finally, the term (\ref{windingmode}) brings only a {\it
finite} renormalisation of the couplings and does not affect their
leading (UV) behaviour  as found on field theory
grounds.

\subsubsection{Heterotic string regularisation of the field theory
result}\label{heterotic_reg}
Here we provide a physical (string)  regularisation of the  
effects of Kaluza-Klein states.
We noticed that the coefficient $\pi/3$ in (\ref{piover3}) arises as a 
result of integrating over $\Gamma$ leading to  a finite (power-like with 
the scale) result, 
unlike the case of field theory where a quadratic divergence appears.
For a physical understanding of this result we 
analyse what string  gives without performing the Poisson
re-summation (eq.\ (\ref{stringthreshold})), since it is not clear
how to interpret (in field theory) a Poisson re-summed Kaluza-Klein 
state of a given level. Thus we do not necessarily require the presence of an
infinite  tower of Kaluza-Klein states. 
From eq.\ (\ref{gag}) and (\ref{stringthreshold}) setting $n_{1,2}=0$ gives
that
\begin{equation}
\Delta_i(n_{1,2}=0)\equiv\Delta _{i}^{KK}
=\frac{\overline{b}_{i}}{4\pi }\int\limits_{-1/2}^{1/2}d\tau
_{1}\int\limits_{\sqrt{1-\tau _{1}^{2}}}^{\infty }d\tau _{2}\frac{1}{\tau
_{2}}\bigg\{ \sum_{m_{1},m_{2}\in \bZ} \exp\bigg[ -\frac{\pi \tau
_{2}}{T_{2} U_2}\vert m_2-U m_{1}\vert^2 \bigg]-1\bigg\}. 
\label{mom} 
\end{equation}
Note the similarity of  the Coleman-Weinberg formula in
field theory, eq.\ (\ref{QFTthresholds1}), with that of (\ref{mom})
integrated over $\Gamma$ instead of $(0,\infty)$. 
We  restrict  the sums over $m_{1},m_{2}$ to a finite number of terms
of mass below some field theory cut-off $\Lambda$ to find, 
from (\ref{mom}) and from the relation between $\ga'$
and the string scale $M_s$ (see below eq.\ \eqref{moduli}),
that\footnote{Here we used that, for $z \ll 1$ we have 
\(
{\displaystyle  
\int_{z}^{\infty }\frac{dt}{t}e^{-t}=-\gamma _{E}-\ln
z-\int_{0}^{z}\frac{dt}{t}(e^{-t}-1)\approx -\gamma_{E}-\ln
z.}
\)}
\begin{equation}
\Delta_{i}^{KK}=
\frac{{\overline{b}_{i}}}{2\pi }\sum_{(m_{1},m_{2})
\not=(0,0)}^{{}}\ln \frac{M_{s}}{M_{m_1,m_2}}
-\frac{{\overline{b}_{i}}}{4\pi }\sum_{(m_{1},m_{2})
\not=(0,0)}^{{}}\int\limits_{-1/2}^{1/2}d\tau
_{1}\int\limits_{0}^{\kappa _{\vec{m}}\sqrt{1-\tau _{1}^{2}}}
\frac{dt}{t}(e^{-t}-1), 
\label{fullsum}
\end{equation}
where $M_{m_1,m_2}$ and $\kappa_{\vec{m}}$ are given by
\begin{equation}
M^2_{m_1,m_2}=\frac{\vert m_2-U m_1\vert^2 }{ (\alpha'  T_2)\, U_2},
\,\,\,\,\,\,\,\,\,
(\alpha' T_2)=R_1 R_2 \sin \theta,\,\,\,\,\,\,\,\,\,\,
\kappa_{\vec{m}}\equiv \pi {\alpha'} M^2_{m_1,m_2}. 
 \end{equation}
We denote by $\tilde{\Delta}_{i}$ the second term 
in (\ref{fullsum}) and 
restrict both sums in (\ref{fullsum}) to terms with
$\kappa_{\vec{m}}\ll 1$, thus 
\begin{equation}
M_{m_1,m_2} \ll M_s
\label{kapall1}
\end{equation}
A necessary (but not sufficient) condition for this to hold true is 
(see  (\ref{kkmass}))
\begin{equation}\label{theta}
M_s R_1 \sin \theta \gg m_1^2,\qquad M_s R_2 \sin \theta \gg m_2^2
\end{equation}
to be compared to (\ref{cond1}).
Eqs.\ (\ref{kapall1}) and (\ref{theta})  impose restrictions not only on the radii but 
also on the shape parameter $\theta$ as well. Indeed, even if the radii are large, a
small angle $\theta$ can render the mass of Kaluza-Klein states very large and 
then (\ref{kapall1}) is not respected. If inequality (\ref{kapall1}) holds, then
\begin{equation}
\Delta _{i}^{KK}(\kappa _{\vec{m}}\leq 1)=\frac{{\overline{b}_{i}}}{2\pi }%
\sum_{(m_{1},m_{2})\not=(0,0)}^{\kappa _{\vec{m}}\leq 1}\ln
\frac{M_{s}}{M_{m_1,m_2}}
-{\tilde{\Delta}}_{i}(\kappa _{\vec{m}}\leq 1), 
\label{stringcutoff}
\end{equation}
which generalises a previous result of \cite{Ghilencea:1999cy}
with $U$ dependence (hidden in $M_{m_!,m_2}$) taken into account.
The term ${\tilde{\Delta}_{i}}$ is negligible in the limit (\ref{kapall1})
and we are only left with  (a finite number of) logarithmic terms. 
These are just those  expected in an effective field theory approach. 
Eq.\ (\ref{stringcutoff})  is the effective field
theory limit of the string thresholds, valid when  the 
energy scale  is below/of the order of the string scale, so that only
a {\it finite number} of modes of mass small relative to this scale
contribute with the usual logarithmic corrections. Comparing
(\ref{stringcutoff}) to (\ref{result}) confirms the power-like
(in scale) regime of the couplings only exists  in the limit of
{\it infinite} number of Kaluza-Klein states. 

The above overlap of the results in effective field theory to those of string
theory in the  case of including only a truncated tower of
Kaluza-Klein modes shows how the string regularise the otherwise (quadratically) 
divergent behaviour of the effective field theory, by restricting
to the case where the physical cut-off of the effective theory
satisfies $\Lambda\leq M_s$. Similar mechanism applies at the level of
vacuum energy when only a truncated tower of Kaluza-Klein states
(below the string scale) is included. In this procedure modular invariance of
the (heterotic) string - essentially the lower limit of integration 
$\sqrt{1-\tau_1^2}$ of (\ref{mom}) - played an important, ``regularisation''
role to ensure a finite result. In a sense winding modes are present
in (\ref{mom}) by setting the lower limit of integration of the fundamental
domain. Finally, note that  (\ref{mom}) (which has $n_{1,2}=0$) and 
subsequent equations correspond to the  orbits with
$\det A=0$ or $A=0$, so $\Delta_i^{KK}$ recovers the (leading) 
part due to (\ref{piover3}) and (\ref{detA=0}). However, $\Delta_i^{KK}$ 
misses the part due to summing over winding modes $n_{1,2}$ in
(\ref{mom})  which essentially would bring an integration limit 
$\tau_2\geq 0$ rather than from $\tau_2\geq \sqrt{1-\tau_1^2}$ as in (\ref{mom}) 
for cases with $\det A=0$. This part was however included by 
the result (\ref{result}) of summing an infinite tower of Kaluza-Klein states.

\subsection{Type I String case}
Comparing the field theory result 
eq.\ (\ref{result}) with the type I string result eq.\ (\ref{t1form}) 
we notice the following.
Firstly, there is a logarithmic contribution $\ln T_2$  present in
both the field and
string theory case, and is traced to the infinite sum of the effects 
of  Kaluza-Klein states. 
Second,  there is no winding mode contribution  in type I string case, this
explains  the absence  of  terms of type (\ref{windingmode}) (manifest
in the heterotic case), and  this again agrees with the field theory
calculation,  where it is absent too, eq.\ (\ref{result}).

The similarity field theory - type I results is not surprising since 
the type I string calculation is in fact just 
a summation over a tower of momentum modes just like in field theory,
but over a  geometry (the annulus $\cA$ 
and M\"obius  $\cM$ strip) different from  that considered in 
deriving (\ref{result}) and this has implications for the
comparison with field theory. Indeed, according to 
\cite{Antoniadis:1999ge} one $\cN =2$ sub-sector  gives
\begin{equation}\label{t1}
\Delta_i^I=\frac{{\overline {b}}_i}{4\pi} \int_{0}^{\infty} 
\frac{dt}{t} \sum_{m_1,m_2\in \bZ}' e^{-\frac{\pi\,t}{T_2 U_2}
\left\vert m_1+ U m_2\right\vert^2}\bigg\vert_{reg}.
\end{equation}
The notable difference to field theory calculation is that after a
Poisson re-summation, the coefficient of $T_2$ in $\Delta_i^I$ is just
zero. This also differs from  (\ref{piover3})
of the heterotic case (where the fundamental domain of the torus
gave a $\pi/3$ coefficient to $T_2$). 
Thus the would-be  UV {\it regularised} quadratic
divergence of the M\"obius strip and annulus just cancel each other
in the closed string channel, and this is marked by the 
symbol ''reg'' in (\ref{t1}) to stress the need for a regularisation
even at string level, for both $\cA$ and $\cM$. The coefficients
these two contributions come with are controlled by the tadpole cancellation
condition \cite{Antoniadis:1999ge}, with no clear equivalent at field
theory level. (This situation is similar to the heterotic case where the
coefficient $\pi/3$ of eq.\ (\ref{piover3}) of the leading contribution
could not be recovered on field theory grounds).
Therefore the two approaches differ by the power-like term which is present in
field theory. For the same reason there is no physical regularisation as
in section \ref{heterotic_reg} of the heterotic case, as a smooth
transition from a field theory to a type I string regime.

\section{Conclusions}\label{conclusions}
Using  a field theory approach in a two torus compactification
we computed the effects of an infinite
tower of Kaluza-Klein states to the gauge couplings, 
and included ``shape'' effects not considered before.
These may give significant effects even in the case $R_2\gg R_1$, 
with UV behaviour different from the pure one dimensional case
and with possible phenomenological implications.

The field theory result obtained using cut-off regularisation and
a summation over the tower of Kaluza Klein states only
is very close to that of the heterotic string.
A comparison of  these two results shows that the logarithmic divergence 
is reproduced precisely as in string theory in the limit $\alpha'\ra
0$.  Further, a quadratic divergence  is also obtained in field theory,
in agreement with the string result in this limit.  
The  full  UV behaviour of the gauge couplings  
may be described in a field theory approach, as due to  Kaluza-Klein states, 
{\it only if a suitable (re)definition of the  regulator} is made.
This re-definition is necessary since the coefficient of 
the UV quadratic divergence of the string in the limit $\alpha'\ra 0$
cannot be recovered on pure field theory grounds. This is because
the exact value of this coefficient is controlled by the symmetries 
of the string not manifest at field theory level (i.e. modular
invariance and thus winding modes' presence).

Within a field theory framework one cannot say that the UV 
finite part of the  result is identical to that in string theory
because the field theory calculation corresponds at (heterotic) string level to
the limit $\alpha'\rightarrow 0$ when additional string 
effects may vanish. We identified  this part 
as due to topological excitations (combined
effect  winding - Poisson re-summed momentum modes) associated
with the extra dimensions considered (world-sheet instantons). This part
 is unrecoverable on field theory grounds.
Within a physical regularisation provided by
the heterotic string, we  showed how the string regularises an 
effective  field theory result of including a finite number of
momentum states, each of them bringing the usual logarithmic contribution.

The role of   modular invariance in ensuring an UV  finite result 
was  discussed. It remains unclear to us the implications (if any) 
of using in string case an  {\it infrared}  regulator 
which does not respect world-sheet  symmetries used in computing the
same string integrals.
 We also stress that while the string result
requires an IR regularisation of the (degenerate orbits) string
correction, the field theory approach of summing {\it massive} KK modes
contribution requires an UV regularisation only. 
This may actually point out a relationship between the 
UV and IR regimes of the theory (enabled at string level by 
winding modes).

In type I case  there are no winding modes' effects, and the finite 
part of the field theory result   is just that of the string
calculation;   however, the issue of a physical regularisation in this case
remains  an open question as it seems to be no smooth  transition
from  the field theory regime of including some (light) Kaluza-Klein states 
below the string scale to the full  string  regime.

\section*{Acknowledgements}
The authors thank L. Dixon, S. F\"orste, J. Louis, F. Quevedo, G.G. Ross,
S. Stieberger and M.\ Walter for helpful discussions. 
\\
The work of D.G.\ was supported by  U.K. PPARC SPG research grant. 
S.G.N.\ was supported in part by the European 
Community Human Potential Programme under contracts 
HPRN--CT--2000--00131 Quantum Spacetime,
HPRN--CT--2000--00148 Physics Across the Present Energy Frontier
and HPRN--CT--2000--00152 Supersymmetry and the Early Universe, 
and priority  grant 1096 of the Deutsche Forschungsgemeinschaft. 

\newpage
\section*{Appendix}
\appendix
\def\theequation{\thesection-\arabic{equation}} 
\section{Field theory calculation of the thresholds effects. (I).}\label{computingJ}
\setcounter{equation}{0}

In this section we compute $\cJ_1$ and $\cJ_2$ of  eqs.\ (\ref{dg1}), (\ref{dg2}).
To compute $\cJ_1$ we introduce the integral
\begin{equation}\label{s1}
\cS  (\xi, \rho)=\int_{\xi}^{\infty} \frac{dt}{t} \sum_{m\in\bZ}' e^{-\pi\,
t\, {m^2}\rho },
\end{equation}
and compute it in the limit  $\xi\ra 0$.  Since
\begin{equation}\label{scaling}
\cS \,(\xi, \rho)= \cS\,(\xi \rho,1),\, 
\end{equation}
we will only  need to evaluate explicitly $\cS\,(\xi,1)$. 
With the above definition  $\cJ_1$  is given by
\begin{equation}\label{s2}
\cJ_1=\cS(\xi,1/(T_2 U_2)),
\end{equation}
To evaluate $\cS(\xi,1)$ we proceed as follows
(a prime on a  sum stands for the absence of $m\not =0$ state):
\begin{eqnarray}
\cS\, (\xi,1)
&=& \int_{\xi}^{1} \frac{dt}{t} \sum_{m\in\bZ}' e^{-\pi\, t\, {m^2} }
+\int_{1}^{\infty} \frac{dt}{t} \sum_{m\in\bZ}' e^{-\pi\,
t\, {m^2} }\label{qq1}\\
&=&\int_{\xi}^{1}\frac{dt}{t}\bigg[-1+t^{-1/2}+ t^{-1/2} 
\sum_{p\in\bZ}' e^{-\frac{\pi}{t}\,
{p^2} }\bigg]+\int_{1}^{\infty} \frac{dt}{t} \sum_{m\in\bZ}' e^{-\pi\,
t\, {m^2} }\label{qq2}\\
&=&\int_{\xi}^{1} \frac{dt}{t} \, \left[-1+\frac{1}{\sqrt t}\right]
+\int_{1}^{1/\xi}  \frac{dt}{\sqrt t} \sum_{p\in\bZ}' e^{-\pi\,
t\, {p^2} }+\int_{1}^{\infty} \frac{dt}{t} \sum_{m\in\bZ}' e^{-\pi\,
t\, {m^2} }\label{seclast}\\
&=&\left[\frac{2}{\sqrt{\xi}}+\ln\xi-2\right]
+\int_{1}^{\infty} {dt}\left[ t^{-1/2}+t^{-1}\right]
\sum_{m\in\bZ}' e^{-\pi\, t\, {m^2} },\label{lastline}\qquad\quad \xi\ll 1
\end{eqnarray}
Under the first integral we first added/subtracted
``-1''  which enabled us further
perform a one-dimensional Poisson re-summation eq.\ (\ref{poisson1}) 
and a subsequent change of variable, $t\ra 1/t$, and removed the regulator
in the second integral of (\ref{seclast}).
The integral in (\ref{lastline}) 
is equal to the  limit  $\cK(\alpha\ra 0)$ where we introduced
\begin{equation}
\cK(\alpha)=\int_{1}^{\infty} {dt}\left[ t^{-1/2-\alpha}+t^{\alpha-1}\right]
\sum_{m\in\bZ}' e^{-\pi\, t\, {m^2} }
=  2 \, \pi^{-\alpha}\, \Gamma(\alpha)\, \zeta(2\alpha) - \frac{1}{\alpha (2
\alpha-1)}.
\label{ll2}
\end{equation}
This is just the  Riemman integral representation of 
$\zeta$ function,  see e.g. \cite{Gradshteyn} (9.513), valid for 
all complex/real $\alpha$. 
The limit $\cK(\alpha\ra 0)$ is well defined and equals
 $\cK(\alpha\ra 0)= -\ln\left(4 \pi e^{- 2 -\gamma}\right)$.
From (\ref{s1}), (\ref{s2}), (\ref{lastline}) we  find 
$\cS(\xi,1)$ and the results used in (\ref{dg1}), (\ref{onedimension})
\begin{eqnarray}\label{lll2}
\cS \,(\xi, 1)&=& \ln \xi+\frac{2}{\sqrt\xi} 
-\ln\left(4 \pi e^{ -\gamma}\right), \quad \xi\ll 1, \\
\nonumber\\
\cJ_1=\cS(\xi/(T_2 U_2),1)&=&
-\ln\bigg[4 \pi e^{-\gamma_E} \, U_2
\frac{T_2}{\xi}\bigg]+2\bigg[\frac{T_2 U_2}{\xi}\bigg]^\frac{1}{2},
\qquad  \xi/(T_2 U_2)\ll 1, \label{JJ1}\\
\nonumber\\
\cS\,(1/\Lambda^2, 1/R^2)&=&-\ln\Big[ 4\pi e^{-\gamma_E}  (\Lambda R)^2
e^{- 2 \Lambda R}\Big],\qquad  \quad \qquad 1/\Lambda\ll R, 
\label{onedim}
\end{eqnarray}

\noindent
To compute $\cJ_2$ of  eq.\ (\ref{dg2}) the  analysis proceeds 
identically. We introduce
\begin{equation}\label{r1}
\cR (\xi, \rho)=\int_{\xi}^{\infty} \frac{dt}{t^{3/2}} 
\sum_{m\in\bZ}' e^{-\pi\, t\, {m^2} \rho}, 
\end{equation}
which we compute in the limit $\xi\ra 0$.
Since 
\begin{equation}\label{sscaling}
\cR \,(\xi, \rho)= \sqrt{\rho} \, \cR\,(\xi \rho,1),  
\end{equation}
we only have to compute $\cR(\xi,1)$. With the above definitions, 
$\cJ_2$ is given by
\begin{equation}\label{r2}
\cJ_2= \sqrt{T_2 U_2}\, \cR(\xi,U_2/T_2).
\end{equation}
We have 
\begin{eqnarray}
\cR\, (\xi,1)
  &=&\int_{\xi}^{1} \frac{dt}{t^{3/2}} 
  \sum_{m\in\bZ}' e^{-\pi\, t\, {m^2} }
  +\int_{1}^{\infty} \frac{dt}{t^{3/2}} \sum_{m\in\bZ}' e^{-\pi\,
t\, {m^2} }\label{sr1}\\
&=&\int_{\xi}^{1}\frac{dt}{t^{3/2}}\bigg[-1+t^{-1/2}+ t^{-1/2} 
\sum_{p\in\bZ}' e^{-\frac{\pi}{t}\,
{p^2} }\bigg]+\int_{1}^{\infty} \frac{dt}{t^{3/2}} 
\sum_{m\in\bZ}' e^{-\pi\, t\, {m^2} }\label{sr2}\\
&=&\int_{\xi}^{1} \frac{dt}{t^{3/2}} \, \left[-1+\frac{1}{\sqrt t}\right]
+\int_{1}^{1/\xi}  {dt} \sum_{p\in\bZ}' e^{-\pi\,
t\, {p^2} }+\int_{1}^{\infty} \frac{dt}{t^{3/2}} \sum_{m\in\bZ}' e^{-\pi\,
t\, {m^2} }\label{sseclast}\\
&=&1+\frac{1}{\xi}-\frac{2}{\sqrt{\xi}}
+\int_{1}^{\infty} {dt}\left[ 1 +t^{-\frac{3}{2}}\right]
\sum_{m\in\bZ}' e^{-\pi\, t\, {m^2} }
=\frac{1}{\xi}-\frac{2}{\sqrt{\xi}}+\frac{\pi}{3},\label{r3}
\quad\xi\ll 1,  
\end{eqnarray}
where the last integral is just $\cK(-1/2)=-1+\pi/3$,
see definition (\ref{ll2}). 
From (\ref{r1}), (\ref{r2}), (\ref{r3}) we  find 
\begin{equation}
\cJ_2= \frac{T_2}{\xi}-2 \bigg[\frac{T_2 U_2}{\xi}\bigg]^\frac{1}{2}
+\frac{\pi}{3} U_2, \qquad \xi U_2/T_2\ll 1\label{JJ2}
\end{equation}
used in eq.\ (\ref{dg2}). For a  generalisation of integrals  (\ref{s1}) and
(\ref{r1}) to cases with arbitrary powers of $t$ in the denominators
of the integrands, see Appendix C in \cite{Ghilencea:2001bv}.

\subsection{(Vanishing) errors in the (regularised) field theory result.}
\label{T2U2conditions}
While  computing $\cJ_{1},\,\cJ_{2},\,\cJ_{3}$ we introduced finite
errors $\delta_{i}$ (i=1,2,3) for each $\cJ_i$, respectively. Their  overall sum
should  vanish in the limit  $\xi \ra 0$ for (\ref{dg1}), (\ref{dg2}),
(\ref{dg4}) to hold true.   The errors arise from (\ref{seclast})
 for $\cJ_1$ (see also (\ref{s2}), (\ref{scaling})), from 
(\ref{sseclast}) for $\cJ_2$  (see also (\ref{r2}), (\ref{sscaling})), 
and from  (\ref{dg3}) for $\cJ_3$. Here we evaluate upper bounds on 
each of them.
The errors are
\begin{eqnarray}
\delta_1 &= & \int_\infty^{1/\ugx} \frac {d t}{t^{1/2}} 
\sum_{m \in \bZ}' e^{- \gp t m^2}, \qquad  
\ugx  \equiv \frac {\gx}{T_2U_2}, 
\label{dj1}
\\[2ex]
\delta_2 & = & U_2 \int_{\bgx}^0 \frac{d t}{t^2} \sum_{m}' 
e^{-\gp m^2/t}, 
\qquad
\bgx \equiv \frac {\gx U_2}{T_2},
 \label{dj2}
\\[2ex]
\delta_3 &=& \sqrt{T_2 U_2} \int_\gx^0 \frac{d t}{t^{3/2}} 
\sum_{m\in \bZ}' \sum_{n \in \bZ}' 
e^{-\gp tU_2 m^2/T_2 - \gp T_2 U_2 n^2/t - 2 \gp i mn U_2}.
\label{dj3}
\end{eqnarray}
To obtain bounds on these errors we use that for $a>0$ 
\equ{
\sum_{m \geq 1} e^{- a m^2} \leq \sum_{m \geq 1} e^{-a m} = 
\frac{e^{-a}}{1 - e^{-a}},
\label{series}
}
With this we find  bounds on $\delta_i$ 
\begin{eqnarray}
\vert \delta_1\vert  & \leq  & \frac{2}{1 - e^{-\gp/ \ugx}}
\int^\infty_{1/\ugx} \frac {d t}{t^{1/2}} e^{- \gp t}, 
\qquad  \ugx \equiv \frac {\gx}{T_2U_2}, 
 \\[2ex]
\vert \delta_2\vert  & \leq & \frac{2 \,U_2}{1-e^{-\pi/\bgx}}
 \int_{1/\bgx}^\infty {d t} 
{e^{-\gp t}}, \quad \qquad \bgx \equiv \frac {\gx U_2}{T_2},  
\\[2ex]
\vert \delta_3 \vert  & \leq & \frac{4 \sqrt{T_2 U_2}}{
1-e^{-\pi (T_2 U_2/\xi)}}
 \int^\gx_0 \frac{d t}{t^{3/2}}  
\frac{e^{-\gp t U_2/T_2}}{1 - e^{- \gp t U_2/T_2}}\, 
e^{- \gp T_2 U_2/t}, 
\end{eqnarray}
where for each $\delta_i$  we further ``relaxed''  the inequalities by 
replacing one factor of their 
integrands   originating from (\ref{series}) 
by the (larger) contribution  in front of the integrals.
For $\delta_3$ we use that (for arbitrary $T_2$, $U_2$, $t$)
\equ{
\frac{e^{-\gp t U_2/T_2}}{1 - e^{-\gp t\, U_2/T_2} } \leq  
\frac {T_2}{\gp t\, U_2}. 
}
The integrals for $\delta_1$ and $\delta_3$ are then further
approximated by the maximal value of their integrands times the
integration range. We  find
\begin{eqnarray}
\vert \delta_1 \vert  & \leq  & \frac{2}{1 - e^{-\gp \, (T_2 U_2)/\xi}}
\left[ \frac 3{2 \gp e} \right]^{\frac 32} \frac {\gx}{T_2 U_2}\label{dd1}, 
\\[2ex]
\vert \delta_2 \vert  & \leq & \frac{2}{1-e^{-\pi \, T_2/(\xi U_2)} }\, \frac{U_2}{\pi}\,
e^{- \gp T_2/(\gx U_2)},\label{dd2}
\\[2ex]
\vert \delta_3 \vert  & \leq & 
\frac{4}{1-e^{-\pi\, (T_2 U_2)/\xi}} \left[\frac{5}{2 \pi e}\right]^{\frac{5}{2}}
\frac{\xi}{T_2 U_2}\frac{1}{\pi U_2^2}.\label{dd3}
\end{eqnarray}
Sufficient, but not necessary conditions for the absolute value of the
overall sum of the errors
to vanish are found from  the requirement each of them vanish
separately (the first two are just those of (\ref{JJ1}) and (\ref{JJ2})). 
\begin{eqnarray}
\vert \delta_1\vert \ll 1 : && \frac{T_2 U_2}{\xi} \gg   1,\label{d1} \\
\vert \delta_2\vert \ll 1: && \frac{T_2 U_2}{\xi} \gg   U_2^2,\label{d2}\\
\vert \delta_3\vert \ll 1: && \frac{T_2 U_2}{\xi}  \gg \frac{1}{U_2^2}.
\label{d3}
\end{eqnarray}
Overall, a sufficient condition is
\begin{equation}\label{overall}
\frac{T_2 U_2}{\xi}\gg \max\left\{ U_2^2, \frac{1}{U_2^2}\right\}.
\end{equation}

\newpage
\section{Field theory calculation of the thresholds effects. (II).}
\label{alternative}
\setcounter{equation}{0}
Here we give a different evaluation of the thresholds,
eq.\ (\ref{QFTthresholds1}) by explicitly relating proper time regularisation
with dimensional or zeta function regularisation. 
The procedure is general, may be applied to other calculations
(e.g. scalar potential in Kaluza Klein models)  and is important
for relating results in different regularisation schemes.
The gauge independent part $\Omega$  of (\ref{QFTthresholds1})
defined  by $\Omega_i\equiv {{\overline {b}}_i}/{(4 \pi)}\, \Omega\,$,  may be
split in two integrals, over $(\xi,1)$ and $(1,\infty)$ to give  (a ``prime''
on a double sum stands for the absence of $(m_1,m_2)=(0,0)$ state) 
\begin{eqnarray}\label{eq1}
\Omega&\equiv&\int_{\xi}^{\infty} \frac{dt}{t}
\sum_{m_{1,2} \in \bZ}' \, e^{-\frac{\pi\,t}{T_2 U_2} |U m_1-m_2|^2}\\
\nonumber\\
&=&\int_{\xi}^{1} \frac{dt}{t} \bigg[ 
\frac{T_2}{t} \sum_{m_{1,2} \in \bZ}' \, e^{- \frac{\pi}{t} 
\frac{T_2}{U_2}\, |U m_1-m_2|^2}
+\frac{T_2}{t}-1\bigg] + \int_{1}^{\infty} \frac{dt}{t}
\sum_{m_{1,2} \in \bZ}' \, e^{- \frac{\pi\,t}{T_2 U_2}\, |U
m_1-m_2|^2}\label{eq2}\\
\nonumber\\
&=& \int_{\xi}^{1}\frac{dt}{t}\left[\frac{T_2}{t}-1\right]+
\int_{1}^{\infty} dt \bigg[ T_2 
\sum_{m_{1,2} \in \bZ}' \, e^{-\pi\,t \, \frac{T_2}{U_2} |U m_1-m_2|^2} 
+\frac{1}{t}\sum_{m_{1,2} \in \bZ}' \, e^{-\frac{\pi\,t}{T_2 U_2}\, 
|U m_1-m_2|^2} \bigg]\label{eq3}
\end{eqnarray}
where we used double Poisson re-summation, eq.\ (\ref{poisson2}). 
The second  integral in (\ref{eq3}) which we denote by $\cL$,  is well defined  
both in the UV and IR and is finite.
All possible  divergences in $\Omega$ 
come from  the first integral in (\ref{eq3}). They 
are due to (Poisson re-summed) ``zero momentum modes''  term $T_2/t^2$
(quadratic in scale),  while $-1/t$ (logarithmic) corresponds to
original Kaluza Klein massless modes.

To compute the second integral of (\ref{eq3}) (i.e. $\cL$) we
consider first the quantity $\cL_\ge$ introduced below, eq.\ (\ref{eqn1}), 
which is well defined in the limit $\ge \ra 0$. 
In this limit $\cL_\ge $ equals $\cL$. 
We have
\begin{eqnarray}
\cL_\ge &\equiv & 
\int_{1}^{\infty} dt\,\, t^\ge \, T_2 \,  \sum_{m_{1,2} \in \bZ}' \, 
e^{-\pi\,t \, \frac{T_2}{U_2} |U m_1-m_2|^2}  
+\int_{1}^{\infty} \frac{dt}{t^{1+\ge}} \sum_{m_{1,2} \in \bZ}' \, 
e^{-\frac{\pi\,t}{T_2 U_2} 
|U m_1-m_2|^2}\label{eqn1}\\
\nonumber\\
&=&\int_{1}^{\infty} dt\,\, t^{ \ge} 
\bigg[\frac{1}{t} - T_2 + \frac{1}{t}\!\!
\sum_{m_{1,2} \in \bZ}' \,  e^{- \frac{\pi}{t}\frac{1}{T_2 U_2}
|U m_1-m_2|^2}\bigg] + \int_{1}^{\infty}\!
\frac{dt}{t^{1+\ge}}\!\! \sum_{m_{1,2} \in \bZ}' \, e^{-\frac{\pi\,t}{T_2 U_2} 
|U m_1-m_2|^2}\label{eqn3}\\
\nonumber\\
&=&\int_{0}^{1} \frac{dt}{t^{1+\ge}}\left[1-\frac{T_2}{t}\right]
+\int_{0}^{\infty}\frac{dt}{t^{1+\ge}}
\sum_{m_{1,2} \in \bZ}' \, e^{-\frac{\pi\,t}{T_2 U_2} |U m_1-m_2|^2}.
\label{cL}
\end{eqnarray}
In (\ref{eqn3}),  we added/subtracted ``$-T_2$'' under the first integral in (\ref{eqn3}) 
 which enabled us to perform a 
``double'' Poisson re-summation (\ref{poisson2}), and isolated
its zero mode, $1/t$. For the same integral a change of
variable is then performed, $t\ra 1/t$ while  $\ge$ presence allows us
to split it into two contributions to reach eq.\ (\ref{cL}).

The second integral in (\ref{cL})  which we denote  by $\cM$ 
is very close to a dimensionally regularised 
version  of Kaluza Klein states' contributions to gauge couplings, see
eq.\ (\ref{QFTthresholds1}). 
We thus related the two regularisation schemes via computing
the finite quantity ($\cL$).  We have
\begin{equation}\label{J}
\cM \equiv
\sum_{m_{1,2}\in \bZ}' \int_{0}^{\infty}\frac{dt}{t^{1+\epsilon}} 
\, e^{-\frac {\pi\,  t}{T_2 U_2} | U m_1 + m_2|^2} = 
\gG(-\ge) \Bigl[ \frac {\gp}{T_2 U_2} \Bigr]^\ge  
\sum_{m_{1,2}\in \bZ}' \frac{1}{|U m_1 +m_2|^{-2\ge}},
\end{equation}
We then use that (for derivation see \cite{Elizalde})
\bea\label{summation}
\sum_{m,n\in \bZ}' \frac{1}{|U m + n|^{2s}} &=&
2 \, \zeta(2s) +
\frac{\sqrt{\pi} \Gamma \left(s-\frac{1}{2}\right)}{\Gamma(s)}
|U_{2}|^{1-2s}\,2\, \zeta(2s-1) 
 \nonumber \\ & + & 
 \frac{8 \pi^{s}}{\Gamma(s)} |U_{2}|^{\frac{1}{2}-s}
\sum_{m=1}^{\infty}\sum_{p=1}^{\infty} \left(\frac{p}{m}\right)^{s-\frac{1}{2}}
\cos\left(2\pi p m U_{1}\right) 
K_{s-\frac{1}{2}}\left(2\pi p m |U_{2}|\right), 
\eea
with $K$ the modified Bessel function and $U=U_1+i\, U_2$.
Notice that the source of possible divergences 
$\gG(-\ge)$ in eq.\ \eqref{J} are canceled by terms of eq.\
\eqref{summation}. Therefore, in the remaining two terms we 
may put $s = - \ge = 0$. 
The sums over the  modified Bessel function 
$K_{-\frac 12}$ can be rewritten in terms of the Dedekind eta-function 
as 
\equ{
\sum_{m=1}^{\infty}\sum_{p=1}^{\infty} \sqrt{\frac{m}{p}}
\cos\left(2\pi p\, m \, U_{1}\right) 
K_{-\frac{1}{2}}\left(2\pi p\, m\, |U_{2}|\right) = 
- \frac 1{4 \sqrt{U_2}} 
\left( 
\ln | \get( U) |^2 + \frac {\gp}6 U_2
\right).
}
We thus find
\begin{equation}
\cM=2 \left[\frac{\pi}{T_2 \, U_2}\right]^{\ge}\Gamma(-\ge)\zeta(-2 \ge)-
\ln\left[\vert\eta(U)\vert^4\right]. 
\end{equation}
Using
\equ{
\gG(- \ge)  = - \frac 1{\ge} - \gamma_E, 
\qquad
x^{\ge} = 1 + \ge \ln |x|, 
\qquad
\gz(- 2 \ge) = - \frac 12 + \ge \ln (2\gp),
} 
we also find the alternative form
\begin{equation}\label{AJ0}
\cM  =\frac{1}{\epsilon}-
\ln\Big[ 4 \pi e^{-\gamma_E}\, T_2 \, U_2\,  |\eta(U)|^4\Big].
\end{equation}
From eqs.\ (\ref{cL}) and (\ref{AJ0}) we find the (finite) result for $\cL$
\begin{equation}
\cL = - \ln\Big[ 4 \pi e^{-\gamma_E} \, e^{- T_2} \,
{T_2} \, U_2\,  \vert \eta(U)\vert^4\Big].
\end{equation}
leading to 
\begin{equation}
\Omega = - \ln\Big[ 4 \pi e^{-\gamma_E} \, e^{-\frac{T_2}{\xi}}\,
\frac{T_2}{\xi} \, U_2\,  \vert \eta(U)\vert^4\Big].
\end{equation}
in agreement with the result of eq.\ (\ref{result}) in the text.

To obtain the type of divergence (quadratic and logarithmic in scale)
due to  terms  $T_2/\xi$ in $\gO$ we used the 
cut-off regularization. With the zeta-function 
or dimensional regularisation procedure this behaviour 
can also be recovered (in (\ref{AJ0}))  if an IR regulator is introduced in addition. 
However, the calculations then become  much more involved. 
(This situation is somewhat similar to the regularization of the integral 
$\int d^4 p\frac 1{p^2}$, which is quadratically divergent 
in the cut-off scheme, but  vanishes with
dimensional regularisation.  This is not in contradiction with 
\eqref{COzeta} that relates these two schemes, since that 
integral is not well-defined. By using an IR regulator 
mass, the dimensional regularised result is divergent and 
is quadratically sensitive to this IR regulator mass.)

\newpage
\section{The link among various regularisation 
schemes}\label{link_regularisation}
\setcounter{equation}{0}

In this section relations among various regularisation schemes and in
particular their link with the Coleman-Weinberg integral used in the
text, eq.\ (\ref{QFTthresholds1}), are presented. Each of
them can in principle be used for the threshold calculations
discussed in this work. 
Let $\gD$ be the Laplacian on the compact two torus with the spectrum 
$ M_{m_1, m_2}$ with $m_1, m_2 \in \bZ$. All relations given in this
appendix do not rely on the precise spectrum of the Laplacian
$\gD$. Let $\Box$ be 
the d'Alembertian of the non-compact 4 dimensional space combined with
the compact torus. The basic quantity that we compute is the trace
$\Tr$ of the d'Alembertian $\Box$  to some (integer) power $\ga$
\begin{equation}
I = \Tr\ {\Box}^{-\ga} = 
\tr \int \frac{d^4 p_4}{(2\gp)^4} (p_4^2 + \gD)^{-\ga}.
\end{equation}
The $\tr$ is over the spectrum of the Laplacian $\gD$ only. Such
traces naturally appear when computing loop corrections. However it 
may require regularisation for to be well-defined. 

In $\gz$-function regularisation a complex number $\gd$ is introduced
to define the operator trace $I$ complex function  
\equ{
I_\gz(\gd) = \Tr\ \Box^{- \ga -\gd} = 
\frac{\gG(\ga + \gd -2)}{16 \gp^2 \gG(\ga)} 
\tr\ \gD^{2- \ga - \gd},
}
where after the second equal sign we have performed the integration
over the non-compact four dimensions. Heat-kernel regularisation
$I_H(s) = \Tr \Box^{-\ga} e^{- s \Box}$ 
is related by a Mellin transform to $\gz$-function regularisation 
\equ{
\frac 1{\gG(\gd)} \int_0^\infty ds\, s^{-1+\gd} I_H(s) = I_\gz(\gd).
}
This result and the Coleman-Weinberg integral representation over 
Schwinger proper time 
\equ{
I_{CW}(\gd) = \frac 1{16 \gp^2 \gG(\ga)} \tr 
\int_0^\infty \frac{dt}{t^{3- \ga -\gd}} e^{- t \gD}
= I_\gz(\gd)
}
follow directly from rescalings of the definition of the
$\gG$-function. The Coleman-Weinberg integral is also often regulated
using a cut-off $\xi$ 
\equ{
I_{CO}(\xi) = \frac 1{16 \gp^2 \gG(\ga)} \tr 
\int_\xi^\infty \frac{dt}{t^{3- \ga}} e^{- t \gD}, 
\qquad
\int_0^\infty d \xi\, \frac {\gd}{\xi^{1-\gd}} I_{CO}(\xi) = I_\gz(\gd).
\label{COzeta}
}
Another option is to use dimensional regularisation on the 
non-compact $ 4 $ dimensions. By integrating over the
$D_4= 4 - 2 \ge_4$ dimensions this almost becomes the $\gz$-function
regularised 
result 
\equ{
I_{D}(\ge_4) = \tr 
 \int \frac{d^{D_4} p_4}{(2\gp)^{D_4}} (p_4^2 + \gD)^{-\ga} 
= (4 \gp)^{\ge_4} I_\gz(\ge_4).
}
Finally using dimensional regularisation for both compact and
non-compact dimensions \cite{GrootNibbelink:2001bx}, 
the regularised expression is  given by 
\equ{
I_D(\ge_4, \ge_5) =  
\int \frac{d^{D_4} p_4}{(2\gp)^{D_4}} 
\int_\ominus \frac{d^{D_5} p_5}{2\gp i} 
 (p_4^2 + p_5^2)^{-\ga} 
\tr\ \frac 1{p_5 - \gD}, 
}
with $D_5 = 1 -2 \ge_5$. This can be expressed as a trace over 
the Laplacian $\gD$ and hence can be rewritten in terms of the
$\gz$-function regularised expression.

\newpage
\section{Poisson re-summation of the  string partition function}
\label{poissonresummation}
\setcounter{equation}{0}

String theory on a n-dimensional torus $\cT^n$ is described by 
the partition function $Z$
obtained by summing over the combined lattice points $\gL_{n,n}$
spanned by the winding modes (the lattice  $\ugL_n = \gL_{n,0}$) and by the 
momentum modes (on the dual lattice $\bgL_n = \gL_{0,n}$). Details on string
partition functions and dualities can be found in \cite{Giveon:1994fu}.
In general the lattice obtained by combining a lattice and a dual lattice
of dimensions $m$ and $n$, respectively is denoted by $\gL_{m,n}$. 
Using the Poisson re-summation formulae 
\equ{
\frac 1{V_{\ugL_n}} 
\sum_{p \in \bgL_{n}} e^{2 \gp i p^T x} = 
\sum_{w \in \ugL_{n}} \gd(x - w), 
\qquad 
\frac 1{V_{\bgL_n}} \sum_{w \in \ugL_{m}} e^{-2 \gp i q^T w} = 
\sum_{p \in \bgL_{m}} \gd(q - p), 
\label{PR}
}
the lattice $\gL_{m,n}$ can be turned into the (``winding-like'') 
lattice $\ugL_{m+n}$  or into the dual (``momentum-like'') 
lattice $\bgL_{m+n}$, respectively. The purpose of this section is
then to show that either of these latter two lattices/descriptions are
equivalent, but this does not remove effects due to 
mixing  between winding and momentum modes (Poisson re-summed or not).
In (\ref{PR}) the volume of the  unit (dual) lattice cell is 
denoted by $V_{\ugL_n}$, $(V_{\bgL_n})$.

In the following we choose  to   apply (\ref{PR}) to the following function 
\equ{
G = \sum_{v \in \gL_{m,n}} e^{- \gp \bigl(
v^T a v + 2 b^T v + c
\bigr)},
\quad
a = \pmtrx{ \ga & \gb \\ \gb^T & \gd }, 
\quad 
b = \pmtrx{ \gm \\ \gn}
\label{stGauss}
}
$G$ is a function of a symmetric matrix $a = a^T$, a vector $b$ and a
scalar $c$. Using the properties of Gaussian integrals, one
obtains the following two representations of the function $G = \uG = \bG$ 
using the properties of Gaussian integrals 
\equ{
\uG =  \frac 1{ \sqrt{\det \gd} V_{\bgL_n}}
\sum_{\uv \in \ugL_{m+n}} e^{- \gp \bigl(
\uv^T \ua\, \uv + 2 \ub^T \uv + \uc \bigr)}, 
\qquad 
\bG = \frac 1{ \sqrt{\det \ga} V_{\ugL_n}}
\sum_{\bv \in \bgL_{m+n}} e^{- \gp \bigl(
\bv^T \ba \bv + 2 \bb^T \bv + \bc \bigr)},
\label{PoissonRes}
}
where we assumed that both $\ga$ and $\gd$ are invertible, with 
\equ{
\arry{rclrclrcl}{
\ua &=& \pmtrx{\ga - \gb \gd\inv \gb^T & - i \gb \gd\inv \\ 
-i \gd\inv \gb^T & \gd\inv}, 
&
\ub &=& \pmtrx{ \gm - \gb \gd\inv \gn \\ - i \gd\inv \gn },
&
\uc &=& c - \gn^T \gd\inv \gn;
\\[2ex]
\ba &=& \pmtrx{ \ga\inv & i \ga\inv \gb \\
i \gb^T \ga\inv & \gd - \gb^T \ga\inv \gb }, 
&
\bb &=& \pmtrx{ i \ga\inv \gm \\ \gn + \gb^T \ga\inv \gm},
&
\bc &=& c - \gm^T \ga\inv \gm.
}
\label{PoissonResPara}
}
In obtaining $\uG$ the momentum modes were ``Poisson re-summed'' to 
leave a sum over
the ``winding-like'' lattice ($\ugL_{m+n}$)  while in $\bG$ the winding modes
were ``Poisson re-summed'' to leave a sum over the 
``momentum-like'' lattice ($\bgL_{m+n}$). 
Note that the off diagonal entries in $\ua$  ``mix'' original winding modes
with Poisson re-summed momentum modes, while the off diagonal entries 
in $\ba$ mix  original momentum modes  with Poisson re-summed winding modes.
Also note that the matrix $\ua$ becomes $\ba$ if we perform the following
substitutions $\ga \ra (\ga - \gb \gd\inv \gb^T)\inv$, 
$\gb \ra - \ga\inv \gb (\gd - \gb^T \ga\inv \gb)\inv$ and 
$\gd \ra (\gd - \gb^T \ga\inv \gb)\inv$. 
  
We can now apply the above formulae in the context of
string theory. The Hamiltonians for the left and right movers on 
the string on a $n$ dimensional torus with metric $g$ and
anti-symmetric tensor $b$ is 
\equ{
H_\pm  = \frac12 P_\pm^T g P_\pm = \frac12 v^T G_\pm v,
\quad \text{with} \quad
G_\pm  =  \pmtrx{ g - b g\inv b &  \frac 12 (\pm1 + b g\inv) \\
\frac 12 (\pm 1 -  g\inv b) & \frac 14 g\inv}
}
derived from the momenta 
\(
P_\pm = \pm w + \frac12 g\inv p - g\inv b w.
\)
The resulting string partition function reads   
\equ{
Z = \sum_{(w\ p)^T \in \gL_{n,n}} e^{2 \gp i(\gt H_+ - \bar\gt H_-)} = 
 \sum_{v \in \gL_{n,n}} e^{
- \gp v^T \bigl( 2 \gt_2 G_- - i \gt \gS \bigr)v, 
}
\qquad 
\gS = \pmtrx{ 0& 1 \\ 1 & 0}.
}
Since $Z$ is of the standard form of eq.\ \eqref{stGauss}, we can apply
the Poisson re-summation formulae \eqref{PoissonRes} with 
$b = 0$ and $c = 0$. Thus  only $\ua$ and $\ba$ have to be computed in this
case using \eqref{PoissonResPara}.  With the notation $\ug = 2 g$
and $\bg = \frac 12 (g - b g\inv b)\inv$, we find for $\ua$ and $\ba$
\equ{
\ua = \frac 1{\gt_2}
\pmtrx{ |\gt|^2  \ug & -( \gt_1 +  i \gt_2 g\inv b)\ug \\[1ex]
-\ug (\gt_1 - i \gt_2 b g\inv)\  & \ug},
\ 
\ba = \frac 1{\gt_2}  
\pmtrx{ \bg & \bg (\gt_1 + i \gt_2 b g\inv ) \\[1ex]
(\gt_1 - i \gt_2 g\inv b)\bg & |\gt|^2 \bg}.
}
This shows that, provided that the appropriate metric is taken into
account, the two descriptions below with either 
Poisson re-summed momentum modes or Poisson re-summed
winding modes respectively, are equivalent:
\equ{
Z = \frac 1{\gt_2^{n/2}} \frac{\sqrt{\det \ug}}{V_{\bgL_n}} 
\sum_{\uv \in \ugL_n} e^{- \gp \uv^T \ua\, \uv} = 
\frac 1{\gt_2^{n/2}} \frac {\sqrt{\det \bg}} {V_{\ugL_n}} 
\sum_{\bv \in \bgL_n} e^{- \gp \bv^T \ba\, \bv}.
}
This equivalence justifies an expectation for some  similarity 
between  {\it regularised} field theory (with  
{\it infinite} towers of momentum states only) and (heterotic) string results.
As already noted, the off diagonal entries in $\ua$  ``mix'' original winding modes
with Poisson re-summed Kaluza-Klein modes, while the off diagonal entries 
in $\ba$ mix  original Kaluza-Klein modes  with Poisson re-summed winding modes.
The off-diagonal term with $b$ does not depend on $\gt_2$ (thus 
unlikely to bring UV effects). Because of the 
anti-symmetry of $b$ we can divide the Poisson 
re-summed and original Kaluza-Klein 
or winding modes into three classes: 1. the zero orbit, which has all 
Poisson re-summed Kaluza-Klein (winding) and original winding
(Kaluza-Klein) modes set to
zero; 2. the degenerate orbit, where the re-summed Kaluza-Klein (winding) and 
original winding (Kaluza-Klein) modes are linearly dependent;  
3. the non-degenerate case. 
 
By applying the formalism developed above to the case of a two torus
with generic shape (as considered in the main text) one finds the
``double'' Poisson re-summation  formula
\begin{equation}\label{poisson2}
\sum_{m_{1,2} \in \bZ} e^{-{\pi\,t}\frac{1}{T_2 U_2} |U m_1-m_2|^2} 
=  \frac{T_2}{t}
\sum_{\tilde m_{1,2} \in \bZ} e^{-\frac{\pi}{t} \frac{T_2}{U_2} |\tilde m_1+U
\tilde m_2|^2}
\end{equation}
where as usual $U_1$ and $U_2$ denote the real and imaginary parts of
the complex shape parameter $U$, respectively.  Also 
\begin{equation}\label{poisson1}
\sum_{n \in \bZ} e^{-\pi A (n-\sigma)^2} = \frac{1}{\sqrt A}
\sum_{\tilde n\in \bZ} e^{-\pi A^{-1} {\tilde n}^2  - 2 i \pi  {\tilde n}  \sigma} 
\end{equation}

\end{document}